\documentclass[10pt,draft]{dis03}
\usepackage{epsf,amsmath}
\usepackage{epsfig}

\begin{document}

\title{Search for Single Top Production in ep Collisions at HERA
\thanks{This talk was given on behalf of the H1 collaboration}}

\author{Andr\'e Sch\"oning\\
ETH Z\"urich, H\"onggerberg \\
CH-8093 Z\"urich, Switzerland\\
E-mail: schoning@mail.desy.de }

\maketitle

\begin{abstract}
\noindent 
In this talk a search for the single production of top quarks
performed at the H1 experiment is presented. 
No evidence for anomalous top production via FCNC processes is found
and limits on the anomalous magnetic coupling constant $\kappa_{tu\gamma}$ are given.
\end{abstract}

\section{Introduction}
Since the first observation 
of an isolated muon event with large 
missing transverse momentum ($p_T^{\rm miss}$) and a high $p_T$ jet
\cite{the_event} by the H1 experiment, 
this ``isolated lepton'' event class has been the subject of 
many studies performed at HERA
\cite{h1_iso_98,h1_iso_03,zeus_iso}.
The main SM process with a charged lepton, a jet and a neutrino
($p_T^{\rm miss}$) in the final state is the production of W~bosons which decay leptonically.
For large transverse momenta of the hadronic final state, $p_T^X>25$~GeV,
six events (1e + 5$\mu$) were observed for an expectation of 3.2 \cite{h1_iso_98} 
in the analysis of the 36.5~pb$^{-1}$ of $e^+p$ data taken from 1994-97.
Recently, the full HERA~I data set has been analysed \cite{h1_iso_03}
combining $e^+p$ data and $e^-p$ data with a total integrated luminosity
which is 118.3~pb$^{-1}$. With an improved SM background suppression,
10 events were observed to be compared to 2.9 expected, confirming the
excess. A similar analysis performed by the ZEUS collaboration
\cite{zeus_iso} does not confirm the excess; their data
were found to be in agreement with the SM. However, in a recent analysis
of the $t \rightarrow b \; \tau^+ \nu_\tau$ channel 2 events were seen to be compared 
to 0.12 expected \cite{zeus_tau}. 

The isolated lepton event class was discussed as a possible signal for the
anomalous production of single top quarks at HERA \cite{theor_top}
with subsequent semi\-leptonic decay: $t \rightarrow b \; \ell^+ \nu$.
The SM expectation for single top quark production at HERA is of the
order of $1$~fb and is negligible. Therefore single top production 
at HERA provides an excellent testing ground for the search for FCNC 
processes beyond the
SM. These processes are predicted for instance by  supersymmetric models, 
 dynamical EWSB\footnote{Electroweak Symmetry Breaking}, extended
 Higgs sectors or other additional symmetries.

In this talk a search for the single production of top quarks
performed at the H1 experiment is presented. 

\subsection{Selection of the leptonic decay \boldmath{$\ t \ \rightarrow \ b \; \ell^+ 
    \nu$}}
The search here presented \cite{top_2002} is based on the selection of
isolated lepton events \cite{iso_2001}. The main preselection cuts are:
a minimum transverse momentum of the electron or muon
$p_T^\ell>10$~GeV, a missing neutrino $p_T^{\rm miss}>10$~GeV, a high
$p_T^X>25$~GeV hadronic final state, 
acoplanarity between the lepton and the hadronic
final state and the isolation of the lepton from jets in the
event $D_{\ell, \rm jet}>1$. 
In addition, specific selection criteria to enrich the single top
signature are applied: the b-jet has to have a minimum $p_T$ of
$25$ ($35$)~GeV in the central (forward\footnote{The direction of the
  proton beam defines the forward region.}) part of the detector;
the transverse mass of the lepton-neutrino system $M_T^{\ell \nu}$ has to exceed
$10$~GeV and the reconstructed charge of the top quark has to be
positive, as HERA is mainly sensitive to the FCNC transition
$u\rightarrow t$. The top selection efficiency was simulated
using the ANOTOP generator, see \cite{iso_2001} and references therein, and found to be 37\% for 
the electron channel and 45\% for the muon channel.
After this selection, H1 sees $5$ events for $1.8\pm0.5$ expected.
By applying a $W$ mass constraint, the top candidate mass can be
calculated after reconstructing the neutrino kinematics. For each
event up to two solutions are possible. 
All calculated masses are compatible with the expected top mass
distribution. 
 
\subsection{Selection of the hadronic decay: \boldmath{$\ t \ \rightarrow \ b \; j \; j$}}
Hadronic top candidates are searched for in the three jet final state
topology. The three jets are ordered in
$p_T^{\rm jet}$ and are required to have more than $40$, $25$, $20$~GeV,
respectively. Only jets in the central and forward part of the detector,
$-0.5 < \eta_{\rm jet} < 2.5$, are considered, with $\eta$ being the pseudorapidity. The total transverse
energy has to fulfill $E_T^{\rm tot}>110$~GeV and events with
identified electrons are rejected to suppress background from
NC deep inelastic scattering.
The top decay is then reconstructed to further suppress the SM background. 
The two jets originating from the $W$ decay are
selected by choosing that jet pairing for which the invariant mass $M_W^{jj}$
is closest to the nominal $W$~mass. This prescription gives the correct
pairing in
$70\%$ of the events.
The reconstructed $W$ mass is then
required to be compatible with the real mass: $70 < M_W < 90~{\rm GeV}$.
To further suppress QCD background, the opening angle $\Theta^\star$ 
of the two $W$~jets reconstructed in the top rest frame has to fulfill
$\cos{\Theta}^\star > -0.5$ (acolinearity) and the top
mass reconstructed from all three jets has to fulfill $150 < M_{jets}
< 210$~GeV. 
The top selection efficiency after all cuts is 27\%.
In total 14 candidate events are found for a SM expectation of
$19.6\pm7.8$. 
The systematic error of about $40\%$ is largely dominated by
a $30\%$ uncertainty on the QCD background prediction.

\section{Results}
\begin{figure}[!thb]
\vspace*{8.0cm}
\begin{center}
\includegraphics{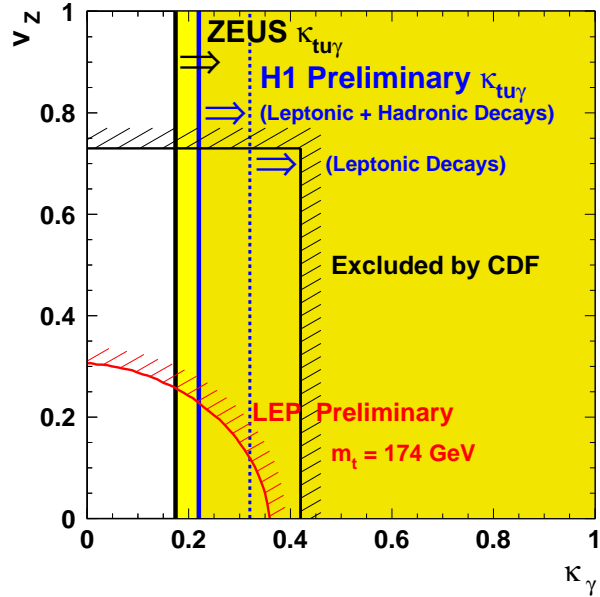}
\caption[*]{Summary plot of experimental limits on the anomalous
  coupling constants $\kappa_{\gamma}$ and $v_Z$. H1 limits are valid
  for the anomalous $\kappa_{tu\gamma}$ coupling only and were
  derived from the leptonic decay only and from the combination with the
  hadronic decay.}
\label{ano_results}
\end{center}
\end{figure}

In the leptonic top decay a slight excess of events is
observed whereas in the hadronic channel a slight deficit is seen.
By combining both channels a  limit at the 95\% confidence level on the top cross
section of $\sigma_{320}(ep\rightarrow etX)<0.43$~pb is obtained
for a cms energy of 320~GeV.
Here a branching ratio of $BR(t\rightarrow
bW)=100\%$ is assumed. 
The hadronic channel alone
gives $\sigma_{320}(ep\rightarrow etX)<0.40$~pb, which is better than
the combination due to
the excess in the lepton channel.
The combined limit can be directly converted into a
limit on the FCNC anomalous magnetic
coupling of the top quark to a $u$-quark and a photon:
$\kappa_{tu\gamma}<0.22$ (95\% CL). 
This limit includes NLO corrections for both the top signal \cite{top_nlo} and the
SM $W$ background cross section \cite{diener}.
Similar results were also
obtained by the ZEUS collaboration \cite{zeus_results}. These
are slightly more constraining than those obtained by H1 due to the absence of
the excess in the leptonic channel. A comparison of all experimental
limits is shown in Fig.~\ref{ano_results}. This shows also limits
obtained at LEP from a search for single top production and limits
from the Tevatron from the study of radiative top decays.
The figure shows that HERA has an unique discovery potential for an
anomalous magnetic coupling of the top quark in a parameter space not
excluded by other experiments. 

\section{Conclusion and Outlook}
Motivated by the excess of events seen in the isolated lepton topology
the H1 experiment has studied the anomalous production of top quarks.
The interpretation that single top production contributes to the isolated
lepton events is not excluded by the results obtained in the
hadronic channel, which suffers from large systematic errors.
Therefore, higher order QCD calculations are vital to improve the top
sensitivity in this channel.
HERA has a unique discovery potential for anomalous top quark production.
A further improvement of the sensitivity is expected with the startup
of HERA~II, which will deliver higher luminosities and larger
event samples in the coming years.

\section*{Acknowledgments} 
I would like to thank my colleagues from H1 for their help in the preparation of
this talk.

\end{document}